\documentstyle[graphicx,12pt]{article}
\begin{document}

\small
\hoffset=-1truecm
\voffset=-2truecm
\title{\bf The evolution of circular loops of a cosmic string with periodic tension}
\author{Leilin Wang \hspace {1cm}Hongbo Cheng\footnote {E-mail address:
hbcheng@sh163.net}\\
Department of Physics, East China University of Science and
Technology,\\ Shanghai 200237, China\\
The Shanghai Key Laboratory of Astrophysics,\\ Shanghai 200234,
China}

\date{}
\maketitle

\begin{abstract}
In this paper the equation of circular loops of cosmic string with
periodic tension is investigated in the Minkowski spacetime and
Robertson-Walker universe respectively. We find that the cosmic
string loops possessing this kind of time-varying tension will
evolve to oscillate instead of collapsing to form a black hole if
their initial radii are not small enough.
\end{abstract}
\vspace{8cm} \hspace{1cm} PACS number(s): 98.80.Cq, 11.27.+d

\newpage

The topological defects such as domain walls, cosmic strings,
monopoles etc. are expected to generate naturally in the early
universe due to phase transitions followed by spontaneously broken
symmetries. The cosmic strings, including their formation,
evolution and observational effects, attracted more attentions of
the physical community in the 1980s and most of 1990s [1, 2]. As
one-dimensional defects at a symmetry breaking phase transition in
the early universe, cosmic strings can arise generally at the end
of an inflationary era within the framework of supersymmetric
grand unified theories [3, 4], which provides us with a potential
window on M theory [5-7]. Cosmic strings could not be the dominant
source of the primordial fluctuations associated with the large
scale structure formation in the universe because of limits on
their tension $G\mu\leq10^{-6}$, but they can still be a secondary
source of fluctuation which can not be neglected [8-10]. The
cosmic strings have also several potentially important
astrophysical features such as gravitational lensing effects
[10-13], gravitational wave background [14-16], early reionization
[17, 18], etc..

A lot of effort has also been contributed to the cosmic string
loops. At any epoch in the history of the universe once the cosmic
strings formed, they are not static and would envolve under their
own tension continuously instead. Within their evolution the
cosmic strings collide and intersect to undergo reconnections
although the strings stretch under the influence of the Hubble
expansion or the environment and the strings lose energy as
gravitational radiation when they oscillate. The results of
reconnections of long strings and large loops is that more and
more small loops will be generated copiously. It is clear that the
cosmic strings exist as string network consisting of long strings
and closed string loops. The cosmic string loops can oscillate
with time rather randomly to become the complicated time-dependent
gravitational sources. Schild et.al. observed and analyzed the
anomalous brightness fluctuation in the multiple-image lens system
Q0957+561A, B which has been investigated for many years [11, 19,
20]. They thought that the phenomena that the system consists of
two quasar images separated by approximately 6 degree are known to
be images of the same quasar not only because of the spectroscopic
match, but also because the images fluctuate in brightness, and
the time delay between fluctuations is always the same. This
effect may be due to lensing by an oscillating loops of cosmic
string between the lensing system and us because cosmic string
loops can provide with a quantitative explanation of such
synchronous variations in two images. In addition the cosmic
string loops can also give rise to the distinct signatures [13,
14, 21]. The contribution of kinks on cosmic string loops to
stochastic background of gravitational waves is estimated [15]. A
large complicated loop of cosmic string fragment and the shape of
small loops are investigated [22]. The number density distribution
of cosmic string loops at any redshift soon after the time of
string formation is derived analytically based on the
Polchinski-Rocha model of loop formation from long strings [23].
Having considered the dynamics of cosmic strings and cosmic string
loops in anisotropic backgrounds, the authors of Ref. [24] show
the imprint on a cosmic string network in the process before the
universe got to isotrope during inflation. In a word once the
cosmic strings formed, they evolve to generate the cosmic string
loops inevitably, so the evolution and fate of cosmic string loops
have attracted more attention. More contributions have been made
to the research in various cases. In the Minkowski spacetime and
the Robertson-Walker universe, the loops will collapse to form
black holes or become a long cosmic string instead of remaining
oscillating loops [25, 26]. In de Sitter worlds only loops with
larger initial radii can survive [25, 27]. In the Kerr-de Sitter
surrounding, around rotating gravitational sources with positive
cosmological constant, a lot of cosmic string loops, including
smaller ones, can evolve to survive when the gravitational source
rotates faster [28]. The evolution of cosmic string loops is also
discussed in the Gauss-Bonnet-de Sitter spacetimes [29]. The
dynamics of cosmic string loops which carry current is
investigated around the Schwarzschild black hole with a repulsive
cosmological constant [30]. It should be pointed out that for all
examples mentioned above the ten tension of cosmic string is
chosen to be constnt.

It is necessary to explore the evolution of cosmic string loops
when the string tension is changeable. It should be emphasized
that the tensions of cosmic strings are constant is just an
assumption. In the cosmological situations a lot of cosmic strings
have time-varying tension. The important issue that the tensions
of cosmic strings can depend on the cosmic time was put forward by
Yamaguchi [31]. Further the string tension $\mu$ can be denoted as
$\mu\propto a^{-3}$, where $a$ is the universe scale factor which
is proportional to $t^{\frac{1}{2}}$ in the radiation-dominated
era and $t^{\frac{2}{3}}$ in the matter-dominated era respectively
[31, 32]. There cosmic strings whose tension depends on the power
of the cosmic time assumed as $\mu\propto t^{q}$ going into the
scaling solution when $q<1$ in the radiation domination and
$q<\frac{2}{3}$ in the matter domination. The cosmological
imprints from cosmic string with changeable tension and
conventional cosmic strings with constant tension could be
different. The cosmic strings with time-varying tension can also
collide and intersect to undergo reconnections. During this
process the loops of this kind of cosmic string emerge. We paid
our efforts to the evolution of cosmic string loops with
changeable tension in the Robertson-walker universe and de Sitter
spacetimes respectively [33, 34]. We find that the cosmic string
loops whose tension depends on some power of the cosmic time
should not collapse to form a black hoe if the power is lower than
a critical value belonging to thee  power's region mentioned above
in Ref. [31, 32], which means that a lot of cosmic string loops
can survive in the universe. The evolution of circular loops of
cosmic string with time-dependent tension is discussed in the BTZ
black hole background [35]. We should point out that although the
cosmic string loops with time-dependent tension could evolve to
exist, the changeable tension as a decreasing function of time is
just assumed to be related to the power of the cosmic time simply.
When the power of time is negative and only when the negative
number is lower than a critical value, these cosmic string loops
will not contract to their Schwarzschild radii. The dependence of
tension on time may be more complicated. The authors of Ref. [36]
introduced z-dependent coupling in Abelian-Higgs vortex string to
put forward that the string tension is periodic subject to the
length of string, which means that the magnitude of tension
oscillates along the string instead of getting bigger and bigger
or smaller and smaller. As the loops evolve to change their size,
the magnitude of tension oscillate and the periodic tension of
string will modify the movement of string loops. According to our
knowledge little contribution is made to investigate this topic.

In our paper we plan to derive the equation of circular loops of
cosmic string with periodic tension in the Minkowski spacetime and
Robertson-walker universe. We wonder how the periodic change of
tension influences on the evolution and fate of the cosmic string
loops. First of all, we search for the equations of circular loops
of cosmic string in the static and flat spacetime and in the
expanding world respectively by means of the Nambu-Goto action
while the string tension is associated with the loop size. We
substitute the string tension expression which shows its
periodicity depending on the loop's radius into the equation
describing the motion of loop. We solve the equations numerically
to study the evolution of loops and the periodicity of tension on
the fate of cosmic string loops. At the end of paper, the
conclusions and discussions are emphasized.

We research on the evolution of cosmic string loops whose tensions
are functions of loop size in the expanding universe. The
Robertson-Walker metric reads,

\begin{equation}
ds^{2}=dt^{2}-R^{2}(t)(dr^{2}+r^{2}d\theta^{2}+r^{2}\sin^{2}\theta
d\varphi^{2})
\end{equation}

\noindent where the scale factor is

\begin{equation}
R(t)=R_{0}t^{\alpha}
\end{equation}

\noindent Here we choose $\alpha=\frac{1}{2}$ for
radiation-dominated era and $\alpha=\frac{2}{3}$ for
matter-dominated era respectively. When the string tension changes
periodically according to the loop evolution, the Nambu-Goto
action for a cosmic string is given by,

\begin{equation}
S=-\int d^{2}\sigma\mu(t)[(\frac{\partial
x}{\partial\sigma^{0}}\cdot\frac{\partial
x}{\partial\sigma^{1}})^{2}-(\frac{\partial
x}{\partial\sigma^{0}})^{2}(\frac{\partial
x}{\partial\sigma^{1}})^{2}]^{\frac{1}{2}}
\end{equation}

\noindent where $\mu$ is the string tension and the function of
loop size. $\sigma^{a}=(t, \varphi)$, $(a=0, 1)$ are time-like and
space-like string coordinates respectively. $x^{\mu}(t,\varphi)$
($\mu,\nu=0,1,2,3$) are the coordinates of the string world sheet
in the spacetime. For simplicity, let us assume that the string
sheet we focus on lies in the hyper-surface
$\theta=\frac{\pi}{2}$, then the spacetime coordinates of the
world-sheet parametrized by $\sigma^{0}=t$, $\sigma^{1}=\varphi$
can be chosen as $x=(t, \varphi)$.

In the case of planar circular loops, the radius of loop is just
associated with the cosmic time like $r=r(t)$, leading the
size-dependent tension changes with the time. According to the
metric (1) and the spacetime coordinates mentioned above, the
Nambu-Goto action (3) is reduced to,

\begin{equation}
S=-\int dtd\varphi\mu R(1-R^{2}\dot{r}^{2})^{\frac{1}{2}}r
\end{equation}

\noindent leading to the following equation of motion for loops,

\begin{eqnarray}
r\ddot{r}+\frac{d\ln\mu}{dt}r\dot{r}(1-R^{2}\dot{r}^{2})
+\frac{\partial\ln\mu}{\partial
r}\frac{r}{R}(1-R^{2}\dot{r}^{2})^{2}\hspace{1cm}\nonumber\\
+\frac{1}{R^{2}}-\dot{r}^{2}+\frac{3\dot{R}}{R}r\dot{r}
-2R\dot{R}r\dot{r}^{3}=0
\end{eqnarray}

\noindent where the dot denotes the differential with respect to
time. Within the frame of Abelian-Higgs model, when the couplings
as periodic functions of length of string were introduced, the
string tension would be periodic in stead of being constant or
decreasing [36]. The approximate expression for tension of this
kind of string is,

\begin{equation}
\mu=\mu_{0}[1+\sin^{2}(\frac{2\pi r}{\Delta})+\sin^{4}(\frac{2\pi
r}{\Delta})]
\end{equation}

\noindent where $\Delta$ is the period. $\mu_{0}$ is a constant.
It is clear that the tension has something to do with the loop
size shown as the perimeter or the radius equivalently because the
circular loop perimeter is $2\pi r$. The magnitude of the string
tension is within a region which can be denoted as
$\mu\in[\mu_{0}, 3\mu_{0}]$ for Eq. (6). When the cosmic string
loops expand or contract, their tensions will become stronger or
weaker alternatively. Now we investigate the evolution of this
kind of cosmic string loops. In the case of Minkowski spacetime,
i.e. $R(t)=constant$, the equation of motion (5) is reduced to,

\begin{equation}
r\ddot{r}+\frac{\frac{2\pi}{\Delta}(1+2\sin^{2}\frac{2\pi
r}{\Delta})\sin\frac{4\pi r}{\Delta}}{1+\sin^{2}\frac{2\pi
r}{\Delta}+\sin^{4}\frac{2\pi
r}{\Delta}}r(1-\dot{r}^{2})-\dot{r}^{2}+1=0
\end{equation}

\noindent Here we set $R(t)=1$. As the equation of circular loops
of cosmic string goes for enough time, once the loop radii
approach a constant meaning $\dot{r}=0$, the equation of motion
becomes,

\begin{equation}
\frac{2\pi(1+2\sin^{2}2\pi r)\sin4\pi r}{1+\sin^{2}2\pi
r+\sin^{4}2\pi r}r+1=0
\end{equation}

\noindent The solutions can be listed partly as $r=0.294, 0.473,
0.767, 0.987$, etc.. It should be pointed out that the constant
radii of loops will lead the constant tension of cosmic string, so
these loops shrink immediately in the Minkowski world [1, 25, 26].
The smaller and smaller size of loops will let the string tension
to be periodic, and the loops start to oscillate. The equation can
be solved numerically. We find that there must exist a critical
value $r_{f}=0.301$ while the period is chosen as $\Delta=1$. When
the initial radius of cosmic string loops $r(t_{0})>r_{f}$ under
$\dot{r}(t_{0})=0$, the loops will enlarge or contract
alternatively or contrarily will collapse to form black holes. The
evolution of radii of circular loops with periodic tension in the
Minkowski spacetime is depicted in Fig. 1. It is shown that not
all cosmic string loops will become black holes in the Minkowski
background like in the case of constant tension of cosmic string.
According to the scale factor (2) as the description of the
Robertson-Walker universe and the periodic string tension (6), the
equation of motion (5) becomes,

\begin{equation}
r\ddot{r}+\frac{\frac{2\pi}{\Delta}(1+2\sin^{2}\frac{2\pi
r}{\Delta})\sin\frac{4\pi r}{\Delta}}{1+\sin^{2}\frac{2\pi
r}{\Delta}+\sin^{4}\frac{2\pi
r}{\Delta}}r(\frac{1}{R_{0}^{2}t^{2\alpha}}-\dot{r}^{2})
+\frac{3\alpha}{t}r\dot{r}-\dot{r}^{2} -2\alpha
R_{0}^{2}t^{2\alpha-1}r\dot{r}^{3}+\frac{1}{R_{0}^{2}t^{2\alpha}}=0
\end{equation}

\noindent At late times $t\longrightarrow\infty$ or equivalently
$R(t)\longrightarrow\infty$, some of terms in Eq. (9) can be
neglected and the equation introduce,

\begin{equation}
r\ddot{r}-\frac{2\pi(1+2\sin^{2}2\pi r)\sin4\pi r}{1+\sin^{2}2\pi
r+\sin^{4}2\pi r}r\dot{r}^{2}-\dot{r}^{2} -2\alpha
R_{0}^{2}t^{2\alpha-1}r\dot{r}^{3}=0
\end{equation}

\noindent We find that there exists a constant solution to Eq.
(10). The cosmic string loops with constant radii possess the
constant tension and will collapse in the Robertson-Walker
universe [1, 25, 26]. The decreasing radii of loops lead their
tension to be periodic and the cosmic string loops are in the
oscillating process. The circular loops of cosmic string are
static just for a moment. In Eq. (10) describing the evolution of
loops of cosmic string with periodic tension at late times the
last term involves $t^{2\alpha-1}$ and $\dot{r}^{3}$. The factor
$t^{2\alpha-1}$ will be larger as the time passes. For Eq. (10)
only the derivative of loop radius with respect to time like
$\dot{r}$ should become weaker and weaker in order to mediate the
increasing factor $t^{2\alpha-1}$. For the loops of cosmic string
whose tension changes periodically, the complicated equation of
motion (8) will also be solved numerically by means of burden
calculation. We find that there also exist the special limits on
the initial size of cosmic string loops in energy era,
$r_{R}=0.299$ for radiation-dominated era and $r_{M}=0.295$ for
matter-dominated era. In each era when the initial radii
$r(t_{0})>r_{R}$ or $r(t_{0})>r_{M}$ under $\dot{r}(t_{0})=0$,
respectively, all cosmic string loops will become larger and
smaller by turns instead of expanding endlessly in the case that
the string tension is a decreasing function of cosmic tme.
Contrarily all loops will collapse to form black holes at last. It
is also shown that the larger cosmic string loops will not
collapse to form black holes in the expanding universe if their
tension is in the oscillating state. In the Robertson-Walker world
the evolutions of radii of circular loops with periodic tension in
the radiation-dominated era and matter-dominated era are plotted
in Fig. 2 and Fig. 3 respectively. The evolution of radius of
cosmic string loop becomes damped oscillation when the time is
sufficiently long in the case of expanding universe. The results
in the Robertson-Walker universe are similar to those in Minkowski
spacetime. The smaller string loops will become black holes and
the other loops will evolve to exist in the vibrant state of loops
size.

There could exist a lot of cosmic string loops in our universe
according to the phenomena [10, 19, 20]. We have investigated the
fate of cosmic string loops with decreasing unceasingly tension in
the Robertson-Walker universe except for constant tension ones
[33, 34]. When the tension becomes weaker fast with time, the
cosmic string loops with this kind of tension will expand without
contracting. Here we discuss the evolution of planar circular
loops of cosmic string with periodic tension in the Minkowski
spacetime and the Robertson-Walker universe respectively. The
so-called periodic tension means that its magnitude changes to be
bigger or smaller by turns. It is found that more cosmic string
loops will evolve instead of collapsing if their initial radii are
not small enough. During their evolution the radii of circular
loops vibrate about a configuration of stable equilibrium, meaning
that the loops become larger and smaller in turn instead of
expanding forever. Our findings indicate that there may exist a
considerable number of loops of cosmic string but they are not too
small. The states of the loops are oscillating.

In this work we find the circular loop equation for a cosmic
string with periodic tension evolving in the hypersurface with
$\theta=\frac{\pi}{2}$ in the Robertson-Walker universe. A
remarkable solution to this equation is that a loop may never
contract to one with a Schwarzschild radius if the tension of
cosmic string is periodic and the loops are not too small. The
evolution of this kind of cosmic string loops is an oscillatory
motion although the motion is not strictly periodic. Therefore
more cosmic string loops whose tensions vary periodically within a
region can evolve to survive in our universe except for the loops
of cosmic string with weaker and weaker tension. Our recent
research generalize completely and significantly the previous
results belonging to the case that string tension is a decreasing
function of cosmic time.

\vspace{3cm}

\noindent\textbf{Acknowledgement}

The authors thank the anonymous Reviewer for helpful suggestions.
This work is supported by NSFC No. 10875043 and is partly
supported by the Shanghai Research Foundation No. 07dz22020.

\newpage
\begin{figure}
\setlength{\belowcaptionskip}{10pt} \centering
  \includegraphics[width=15cm]{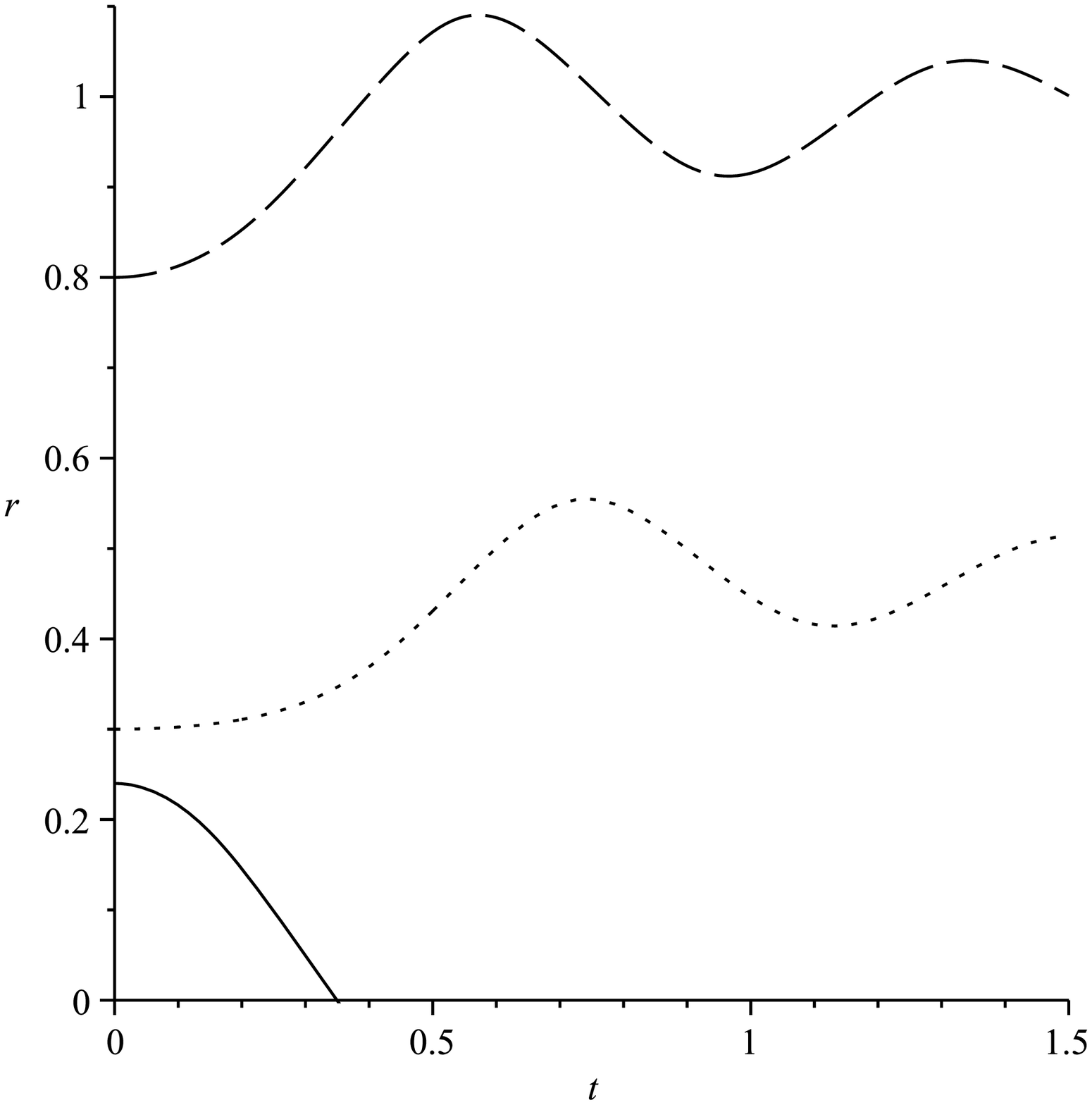}
  \caption{The solid, dotted and dashed curves of $r(t)$, the radii of circular loops
  as functions of cosmic time with initial values $r(t_{0})=0.24, 0.31, 0.8$ respectively
  and $\dot{r}(t_{0})=0$ in the Minkowski spacetime.}
\end{figure}

\newpage
\begin{figure}
\setlength{\belowcaptionskip}{10pt} \centering
  \includegraphics[width=15cm]{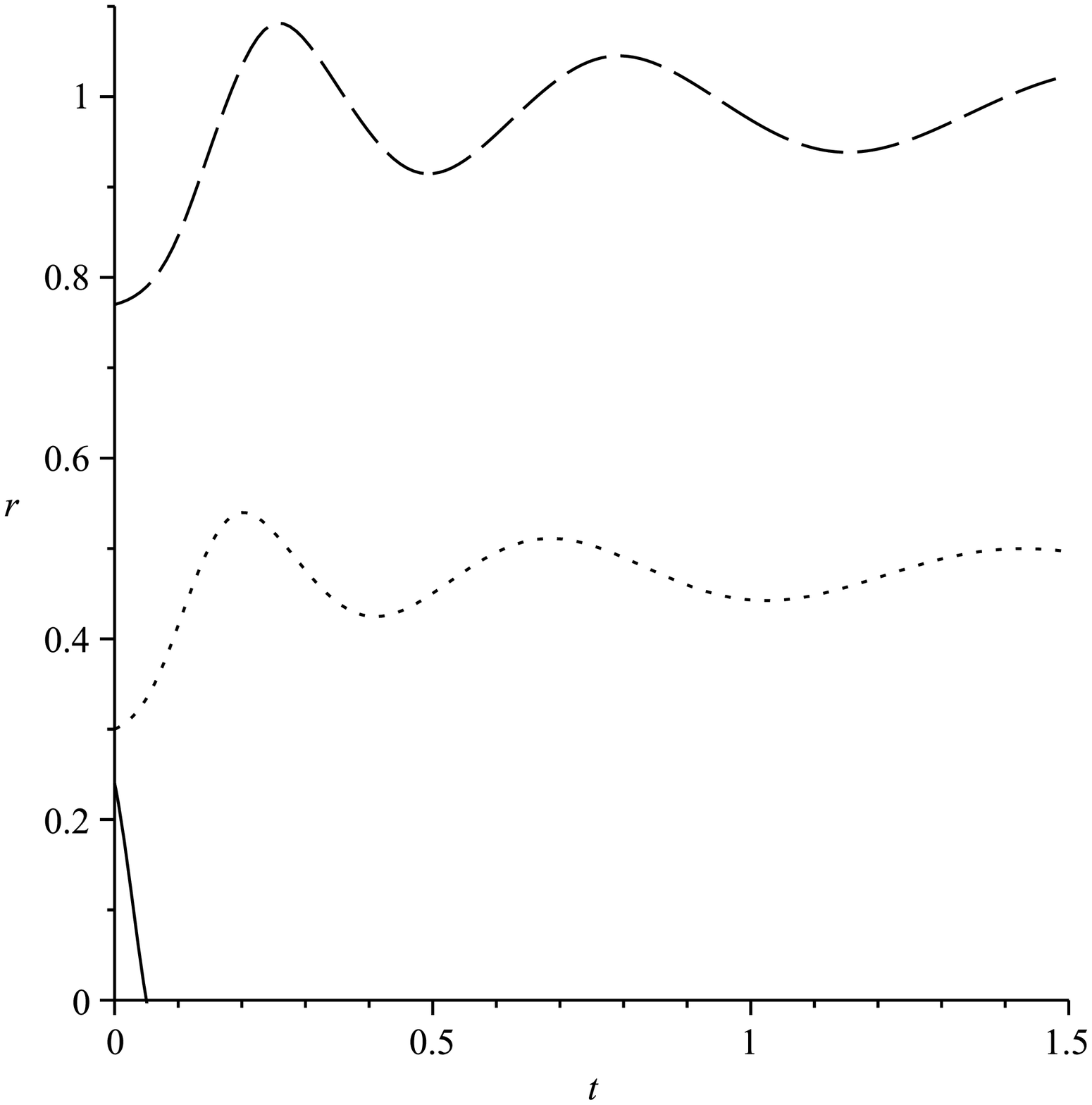}
  \caption{The solid, dotted and dashed curves of $r(t)$, the radii of circular loops
  as functions of cosmic time with initial values $r(t_{0})=0.24, 0.3, 0.76$ respectively
  and $\dot{r}(t_{0})=0$ in the radiation-dominated era.}
\end{figure}

\newpage
\begin{figure}
\setlength{\belowcaptionskip}{10pt} \centering
  \includegraphics[width=15cm]{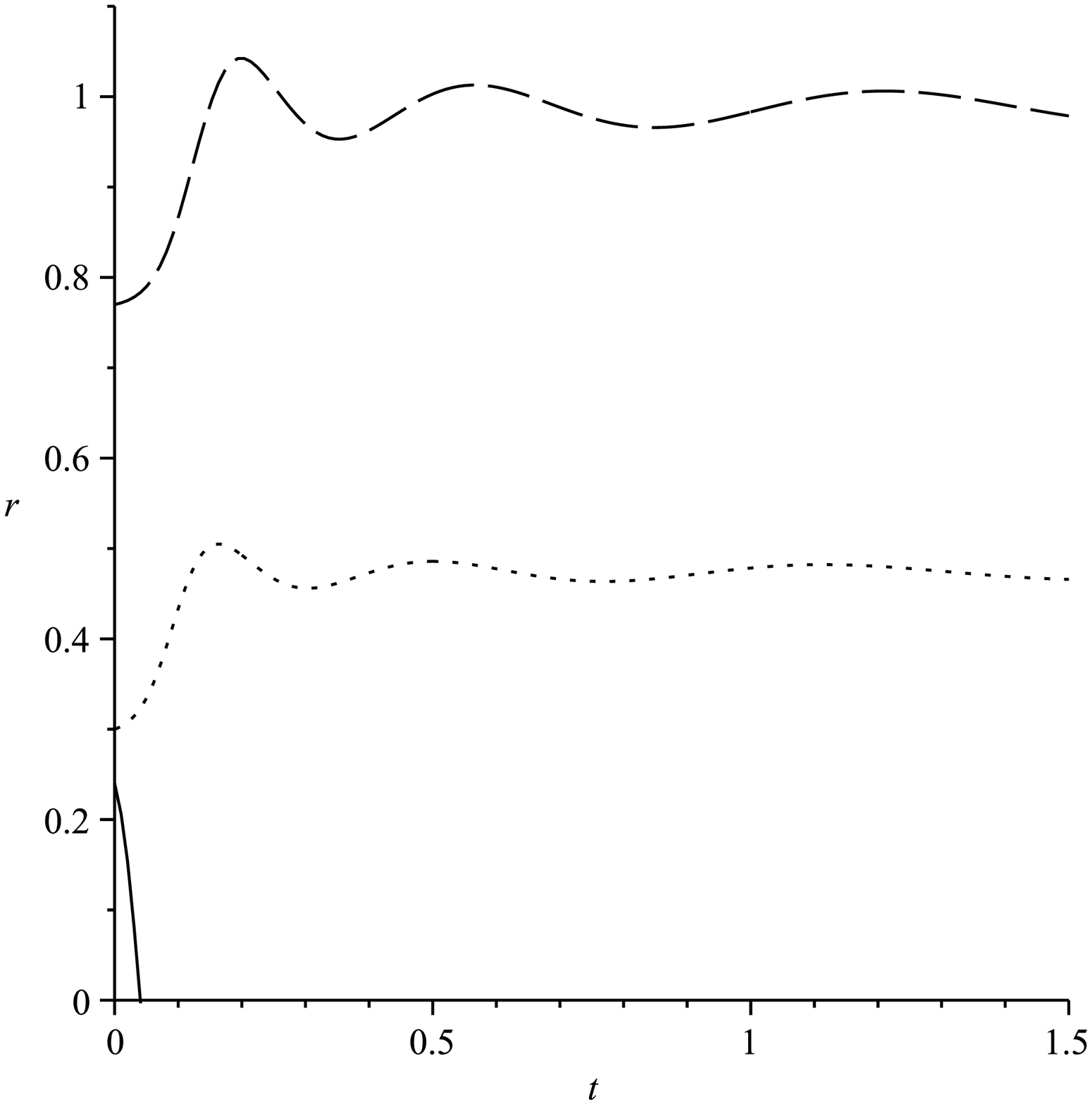}
  \caption{The solid, dotted and dashed curves of $r(t)$, the radii of circular loops
  as functions of cosmic time with initial values $r(t_{0})=0.24, 0.3, 0.76$ respectively
  and $\dot{r}(t_{0})=0$ in the matter-dominated era.}
\end{figure}

\end{document}